\documentclass{llncs}
\usepackage[utf8]{inputenc}
\usepackage{amsmath}
\usepackage{amsfonts}
\usepackage{amssymb}
\usepackage{booktabs}
\usepackage{makecell}
\usepackage{multirow}
\usepackage{soul}
\usepackage{tikz}
\usepackage{todonotes}

\usepackage[hidelinks]{hyperref}

\newcommand*\circled[1]{\tikz[baseline=(char.base)]{
            \node[shape=circle,draw,inner sep=1pt, color=white,fill=black] (char) {#1};}}

\begin{document}

\title{Trust Anchors in Software Defined Networks}	

\author{Nicolae Paladi\inst{1} \and
Linus Karlsson\inst{2} \and
Khalid Elbashir\inst{3}}

\institute{RISE SICS \\ \email{nicolae.paladi@ri.se} \and
Lund University \\ \email{linus.karlsson@eit.lth.se} \and
KTH - Royal Institute of Technology \\ \email{elbashir@kth.se}}

\maketitle

\begin{abstract}
Advances in software virtualization and network processing lead to increasing network softwarization.
Software network elements running on commodity platforms replace or complement hardware components in cloud and mobile network infrastructure.
However, such commodity platforms have a large attack surface and often lack granular control and tight integration of the underlying hardware and software stack.
Often, software network elements are either themselves vulnerable to software attacks or can be compromised through the bloated trusted computing base.
To address this, we protect the core security assets of network elements - authentication credentials and cryptographic context - by provisioning them to and maintaining them exclusively in isolated execution environments.
We complement this with a secure and scalable mechanism to enroll network elements into software defined networks.
Our evaluation results show a negligible impact on run-time performance and only a moderate performance impact at the deployment stage.
\end{abstract}

\section{Introduction}
\label{sec:intro}
Software Defined Networking (SDN) is a widely used approach to operate network infrastructure in virtualized environments.
Separation of forwarding and control logic, a core idea of this model, is often realized by software network elements in a virtualized network infrastructure deployed on commodity hardware.
However, by departing from hardware network elements with tightly couped software and hardware often provided by the same vendor~\cite{etsinfv:2013},
SDN broke previous assumptions, outdated best-practices and introduced new vulnerabilities~\cite{paladi:2015,paladi:2017c}.
Scott-Hayward et al. outlined a series of attack vectors that can lead to unauthorized access, data leakage or modification, malicious applications on the network, configuration issues, and a wider collection of system-level security vulnerabilities~\cite{scott:2015}.
This concern applies to both the data plane and the application plane in SDN deployments.
On the data plane, related literature describes both potential attacks on SDN in case of a virtual switch compromise~\cite{antikainen:2014}, partly demonstrated in~\cite{thimmaraju:2017}.
Malicious applications deployed on the SDN infrastructure are a particular concern in virtualized environments. 
They affect network security both directly (by intercepting or modifying traffic), or indirectly through horizontal attacks aimed to leak authentication credentials and encryption keys~\cite{itu-t:2016}.

Earlier research addressed SDN security through additional services~\cite{porras:2012,shin:2013,hu:2014}, formal verification~\cite{ball:2014} and isolated execution using Intel Software Guard Extensions (SGX)~\cite{shih:2016,paladi:2016b,kim:2017,paladi:2017b}, and most popular network element implementation support communication over transport layer security (TLS)~\cite{dierks:2008}.
Despite these efforts, the confidentiality and integrity of authentication credentials of network elements in SDN remain unaddressed.
In particular, the existing approaches to provision authentication credentials to network elements in SDN are either plain insecure or both insecure and unscalable, requiring manual steps\footnote{Indeed, the Open vSwitch manual contains phrases as 
``Write the fingerprint down on a slip of paper and copy sc-req.pem to the machine that contains the PKI structure''}~\cite{ovs_ssl}.
Moreover, credentials provisioned to network elements in virtualized environments are often stored in plaintext on the file system.
Adversaries exploiting vulnerabilities in process and virtualization isolation can access authentication credentials to perform network attacks or impersonate network elements.
In this paper, we address two complementary questions:
(1) \textit{How can authentication credentials be securely provisioned to software network elements in SDN deployments?}
and 
(2) \textit{How can the TLS context of virtual switches be protected on compromised hosts?}

\subsection{Contributions}
\label{subsec:contrib}
In this work, we present the following contributions:
\begin{itemize}
	\item A secure, practical, and scalable mechanism to provision authentication credentials and bootstrap communication between software network elements.
	\item TLSonSGX\footnote{Source code available: \url{https://github.com/TLSonSGX/TLSonSGX}}, a library allowing to maintain authentication credentials and the TLS context exclusively in isolated execution environments. 
	\item A novel approach to restricting the availability of authentication credentials for SDN components to hosts with an attested trusted computing base.
	\item A first thorough analysis of the performance trade-offs of deploying components of network elements in SGX enclaves.
\end{itemize}

\subsection{Structure}
\label{subsec:structure}
The remainder of this paper is structured as follows.
We present the system model and threat model in~\S\ref{sec:sys-threat-mod}.
Next, we describe the proposed solution in~\S\ref{sec:arch} and its implementation in \S\ref{sec:implem}.
We evaluate the approach in~\S\ref{sec:eval}, discuss the related work in~\S\ref{sec:related-work}, outline limitations and future work in~\S\ref{sec:future_work} and conclude in~\S\ref{sec:conclusion}.

\section{System and Threat Model}
\label{sec:sys-threat-mod}
We consider an SDN infrastructure deployed on commodity platforms in a distributed system, such as in a cloud platform or a mobile communications network.
The infrastructure is managed by the \textit{administrators} of a \textit{network operator}.
Physical access to the platforms is restricted and auditable.

\paragraph{System model} 
Administrators use \textit{orchestrators} to manage network infrastructure, software components and network services~\cite{etsinfv:2013}.
They deploy network elements on the \textit{data plane}, \textit{control plane} and \textit{application plane}. 
The \textit{data plane} consists of hardware or software switches (e.g. Open vSwitch~\cite{pfaff:2015}) and communication links between them.
The \textit{control plane} consists of a logically centralized \textit{network controller} (e.g. ONOS~\cite{berde:2014}, Floodlight~\cite{f:2016}).
The network controller manages software switches through protocols such as OpenFlow~\cite{mckeown:2008} (to add or remove flows) or OVSDB~\cite{pfaff:2013} (to create ports and tunnels); 
it manages hardware switches through OpenFlow (if supported) or other interfaces, such as NETCONF~\cite{enns:2011}.
The \textit{application plane} comprises \textit{network functions} that implement services such as traffic engineering, monitoring, or caching.
A \textit{Virtual Network Function} (VNF) is a virtualisation of a network function~\cite{etsinfv:2013}.
Orchestrators deploy VNFs upon request from the network controller or a tenant. 
The network controller configures flows and steers traffic to the network functions.

Network elements on the data-, control-, and application planes communicate over two application programming interfaces (APIs).
The controller communicates with data plane elements over the \textit{southbound} API, commonly Openflow~\cite{mckeown:2008,sherwood:2010,bifulco:2016} and with application plane elements over the \textit{northbound} API.

At deployment time, the orchestrator provisions TLS certificates to network elements during the \textit{enrollment} process.
Furthermore, to protect the data within the SDN deployment, the network controller enforces communication over TLS with mutual authentication on both southbound and northbound APIs.

\paragraph{Threat model} Similar to earlier work on SDN security threats \cite{kreutz:2013,paladi:2015}, we assume physical security of the platforms underlying the SDN infrastructure and correct implementation of cryptographic algorithms and communication security protocols, such as TLS~\cite{dierks:2008}. 
The adversary has the capabilities of a system administrator with remote access to commodity platforms in the SDN infrastructure.
The adversary can intercept, drop and modify packets on the southbound and northbound interfaces.
Furthermore, the adversary can run arbitrary network elements in the SDN deployment and elsewhere~\cite{etsinfv:2013}. 
The adversary can read the memory of the commodity platforms, exploit vulnerabilities in network elements on the data- and application planes, and circumvent virtualization isolation~\cite{antikainen:2014}.

\section{Solution space}
\label{sec:arch}
We next present the approach for provisioning and protecting authentication credentials on the data and application planes of SDN deployments.
We first introduce three building blocks to create trust anchors in SDN deployments: 
Software Guard Extensions (SGX), Trusted Platform Module (TPM) and Integrity Measurement Architecture (IMA).

\subsection{Trust anchors}
\label{subsec:trust-anchors}
We use SGX enclaves~\cite{anati:2013,mckeen:2013,xing:2016,mckeen:2016} to create trusted execution environments (TEEs) during operating system execution.
We use the TEEs to store authentication credentials and execute cryptographic operations for network elements.
SGX enclaves rely on a trusted computing base (TCB) of code and data loaded at enclave creation time, processor firmware and processor hardware.
Program execution within an enclave is transparent to the underlying operating system and other mutually distrusting enclaves on the platform.
Enclaves operate in a dedicated memory area called the Enclave Page Cache, a range of DRAM that cannot be accessed by system software or peripherals~\cite{mckeen:2013,intel:2017}.
The CPU firmware and hardware are the root of trust of an enclave;
it prevents access to the enclave's memory by the operating system and other enclaves.
Remote attestation~\cite{coker:2011} allows an enclave to provide integrity guarantees of its contents~\cite{anati:2013}.

We use TPMs to store platform integrity measurements collected during boot, and attest the integrity of platforms hosting the SDN infrastructure.
A TPM is a discrete component on the platform motherboard and its state is distinct from the state of the platform.
TPMs provide secure non-volatile storage, cryptographic key generation and use, sealed storage and support (remote) attestation~\cite{tpm:1.2}.
TPMs assume platform integrity by identifying and reporting the platform state that comprises the hardware and software components~\cite{nyman:2014}.
In this context, \textit{trust} is based on the conjecture that a certain behaviour can be expected based on the reported platform state~\cite{paladi:2017c}.
TPMs can prove the association between a cryptographically verifiable identity and the host platform~\cite{tpm:1.2,tpm:2.0}.

We use Linux IMA to measure the integrity of the TCB.
Linux IMA measures a predefined set of files on the system by hashing their contents and storing the values in a measurement list;
it can be configured to detect modifications of files at runtime.
To guarantee the integrity of the measurement list, its trust can be rooted in the TPM.
The system's trustworthiness can be assessed by a remote appraiser by comparing the measurement list to an expected configuration~\cite{coker:2011}.
We utilize IMA to collect measurements of the network elements on the platform.
During the remote attestation of the platform, we use the measurement list to verify the integrity - and implicitly the trustworthiness - of network elements.

\subsection{Data plane}
\label{subsec:data-plane-trust}
At cloud platform deployment time, an orchestrator deploys and runs virtual switches on the underlying compute resources.
To enable network connectivity, the orchestrator instructs virtual switches to add (or delete) ports whenever virtualized execution environments are instantiated or torn down.

For a secure deployment, the administrator must ensure both a secure installation of hardware and software, as well as provision the correct initial configuration of the virtual switch instances in the cloud infrastructure.
In turn, secure generation of keys and provisioning of certificates is a precondition to ensuring security of the initial deployment configuration.
Furthermore, ensuring the integrity of virtual switch binaries and configurations is a precondition for ensuring the run-time security of the deployed instances. 

\begin{figure}[t] 
\centering
\includegraphics[width=0.75\textwidth]{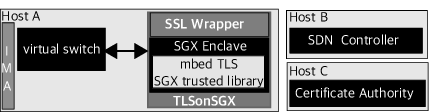}
\caption{TLSonSGX System Design}
\label{fig:tlsonsgx}
\end{figure}

We address this with a new library, \textbf{TLSonSGX}, that enables virtual switches to use a cryptographic library running in a TEE (see Figure \ref{fig:tlsonsgx}).
TLSonSGX provides an abstraction layer and a wrapper around the cryptographic library deployed in a TEE, allowing to easily substitute the implementation depending on performance, functionality and licensing aspects.
Following this approach, TLS sessions originate and terminate within the TEE and the generated keys and certificates are confined to the TEE,
ensuring the confidentiality and integrity of core assets, such as generated keys, certificates and TLS context, even in the event of a host compromise.
This, in combination with an infrastructure monitoring system and a file integrity subsystem (such as Linux IMA), prevents the adversary from impersonating data plane network elements~\cite{thimmaraju:2017} and from enrolling additional network elements into the infrastructure.

Secure provisioning of authentication certificates is challenging, especially at scale, and depends on the capability to establish a secure communication channel between the certificate authority (CA) and the target component.
Several vendor-specific solutions exist~\cite{mckeen:2013,jain:2016}.
To support the deployment, we introduced a CA with extended functionality to sign certificates for the virtual switches and the SDN controller.
CA certificates are provisioned to the virtual switches and the SDN controller in the deployment and are subsequently used for mutual authentication.
Beyond secure certificate provisioning, the extended CA verifies the integrity of the virtual switches before signing their certificates.
We leverage the remote attestation capability provided by the TPM to verify the TCB integrity on the host platform.
The TPM is in this protocol the root of trust that stores and provides a signed quote of the integrity measurements of the virtual switch binary and ancillary libraries, collected by IMA.

\subsection{Application plane}
\label{subsec:application-plane-trust}
Network elements on the application layer, such as VNFs, must be authenticated and integrity verified prior to enrollment into the SDN infrastructure.
As the controller requires mutual authentication with all its clients, this ensures that only trustworthy VNFs can communicate with the controller.
Similar to the approach above, the TPM is used as a root of trust.

We use SGX enclaves to ensure integrity and confidentiality of the authentication credentials for enrolled VNFs.
Storing the credentials in SGX enclaves reduces the attack surface to the enclave TCB and offers an additional layer of protection even in the case of a breach of the platform TCB.
We discuss the limitations of this approach in Section~\ref{subsec:attacks-on-tls-sgx}.

\begin{figure}[t] 
	\centering
	\begin{tikzpicture}[scale=1, every node/.style={scale=1}, node distance=2cm, font=\sffamily]
		\definecolor{bluebase}{RGB}{70,130,180}
		\colorlet{remotecolor}{bluebase!20}
		\colorlet{containercolor}{bluebase!40}
		\colorlet{vnfcolor}{bluebase!60}
		\colorlet{vmcolor}{bluebase!100}
		\colorlet{iascolor}{bluebase!80}
		\colorlet{enclavedcolor}{gray!80}
		\definecolor{enclavevcolor}{RGB}{50,205,50}
		\tikzset{cnc/.style={shape=circle,draw,inner sep=1pt, color=white,fill=black}}
		\tikzset{cn/.style={cnc,above}}
		\tikzset{cnr/.style={cn,right}}
		\tikzset{cnl/.style={cn,left}}
		\tikzset{cnb/.style={cn,below}}
		\tikzset{sa/.style={->,thick}}
		
		\node[minimum width=7cm,minimum height=2.4cm,fill=remotecolor,draw] (container host) at (0, 0) {};
		\node[below right] at (container host.north west) {\small container host};
		
		\node[below left=0.1cm and 0.1cm,minimum width=3cm,minimum height=2.2cm,fill=containercolor,draw] (container) at (container host.north east) {};
		\node[below=-0.05cm] at (container.north) {\small container};

		\node[above right=0.1cm and 0.18cm,minimum width=2.6cm,minimum height=1.75cm,fill=vnfcolor,draw] (vnf) at (container.south west) {};
		\node[below=-0.05cm] at (vnf.north) {\small application};
		
		\node[above=0.1cm,minimum width=1.5cm,minimum height=1.3cm,fill=enclavevcolor,draw] (enclave) at (vnf.south) {};
		\node[below] at (enclave.north) {\small enclave};

			
		\node[right=3cm of enclave,draw,align=center,fill=vmcolor,minimum height=1.3cm] (vm) {(Extended)\\ Certificate\\ Authority};
				
		\node[above right=0.1cm and 0.1cm,minimum width=1cm,minimum height=0.4cm,fill=vnfcolor,draw] (tpm) at (container host.south west) {TPM};
		
		\node[right=1cm of tpm,minimum width=1cm,minimum height=0.4cm,fill=vnfcolor,draw] (ima) {IMA};
		
		\node[above right=-0.1cm and 0.1cm,minimum width=3.2cm,minimum height=0.5cm,fill=vnfcolor,draw,outer sep=0mm] (agent) at (container host.west) {attestation agent};
		\draw[->,thick] (ima.west) -- (tpm.east);

		\coordinate (enclavesu) at ([yshift=5mm]enclave.west);
		\coordinate (enclavesd) at ([yshift=2.5mm]enclave.west);
		
		\draw[sa] ([yshift=5mm]enclave.east) -- node[cn] {1} ([yshift=5mm]vm.west);
		\draw[sa] ([yshift=2.5mm]vm.west) -- node[cnc] {2} ([yshift=2.5mm]enclave.east);
		\draw[sa] (enclavesu) -- node[cn] {3} (enclavesu -| agent.east);
		\draw[sa] (tpm.north) -- node[cnr] {4} (tpm.north |- agent.south);
		\draw[sa] (ima.north) -- node[cnl] {4} (ima.north |- agent.south);
		\draw[sa] (enclavesd -| agent.east) -- node[cnb] {5} (enclavesd);
		\node[cnr] at (enclave) {6};
		\draw[sa] ([yshift=-2.5mm]enclave.east) -- node[cnc] {7} ([yshift=-2.5mm]vm.west);
		\node[below=0.2cm of vm,cn] {8};
		\draw[sa] ([yshift=-5mm]vm.west) -- node[cnb] {9} ([yshift=-5mm]enclave.east);
	\end{tikzpicture}
	\caption{Enrollment steps in the application layer.}
	\label{fig:apparchitecture}
\end{figure}
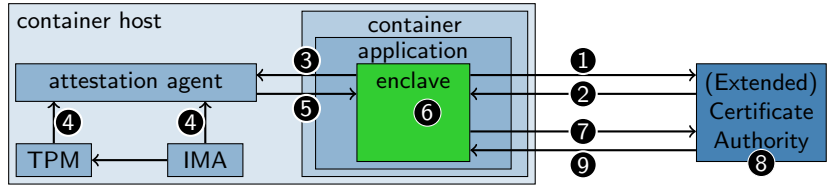

We next provide an overview of the proposed solution (see Figure~\ref{fig:apparchitecture}).
The extended certificate authority (CA) introduced above determines whether or not a VNF configuration is valid, by matching against a list of known good configurations.
If a configuration is valid, the CA can also sign certificates.
This component can be collocated with the network elements in the deployment, or be deployed and operated by a third party.
We assume that the CA root certificate is provided to the SDN controller during initial setup.

At the start of the enrollment protocol, the orchestrator launches an execution environment (such as a bare-metal host, virtual machine or container) with TPM and IMA support.
Together, these two mechanisms record both the software and hardware configuration in a measurement log, including the TCB of the VNFs.
The measurement log is anchored in the TPM located of the host, allowing the use of the TPM's remote attestation functionality.
Note that both a native and a virtualized TPM can be used in this case.

Similar to \cite{zhu:2017} we use an \emph{attestation agent} running on the container host.
This agent proxies the communication between the container and the TPM and IMA.
We propose a solution where the attestation agent is only accessible from the container running on the same host.
This prevents direct communication between the attestation agent and the CA.
To prevent cuckoo attacks \cite{parno:2008}, the communication passes through the container application and the enclave and ensures that the enclave is running on the same host.

The enrollment phase consists of the following steps (see Figure~\ref{fig:apparchitecture}):
Upon initialization of the container and application, the latter requests a nonce from the CA \circled{1},\circled{2}.
Next, the application requests from the attestation agent a quote for the given nonce, together with the IMA measurement list \circled{3}.
The agent communicates with the TPM and the IMA to retrieve the data \circled{4}, and returns the data to the application \circled{5}.
The enclave generates a new private key and a certificate signing request (CSR) and stores it in the SGX enclave \circled{6}.
The application sends the quote, measurement list, and the CSR to the CA \circled{7}, that verifies the message~\circled{8}.
As the measurement list covers both the host system and the container TCB, the integrity of the host and target containers can be validated.
If the measurement values match known good configurations, the CA signs the CSR and returns the signed certificate to the enclave \circled{9}.
At this point, the VNF can establish a secure TLS connection with the SDN controller.
The proposed solution ensures that only trustworthy VNFs receive valid certificates and can be enrolled in the SDN infrastructure.

\section{Implementation}
\label{sec:implem}
To facilitate adoption and obtain reproducible results, we implemented the proposed solution using common open-source libraries and execution isolation features available on commodity platforms.
We used Open vSwitch (OvS), a popular software switch implementation and the Ryu and Floodlight SDN controllers, mainly due to their popularity and simple configuration.
In the remainder of this section, we first describe the implementation of TLSonSGX on the data plane.
Next, we describe the security mechanisms deployed on the application plane.

\subsection{TLSonSGX}
\label{subsec:tlsonsgx-impl}
The SGX programming model requires that applications deployed in SGX enclaves have an external component that can be called by other processes running on the operating system, and that in turn maps such calls to software in the enclave.
This external component is not part of the enclave and its integrity cannot be attested using the SGX integrity attestation mechanisms, thus is is considered \textit{untrusted};
in contrast, the code running in the enclave is considered \textit{trusted} once its integrity has been attested.
Following the SGX programming model, the untrusted code portion of the TLSonSGX library is a wrapper that maps OpenSSL external methods (used by Open vSwitch) internally into enclave calls (ECALLs).
The trusted portion of the code, contained within the SGX enclave, implements the ECALLs by utilizing the SGX trusted TLS library.
Support for TLS libraries in SGX varies and evolves continuously; 
we have chosen the mbed~TLS~\cite{mbedtls_sgx} library considering its sufficient support for SGX enclaves.

Considering that authentication keys and certificates are confined to the enclave, we modified OvS to use only a limited set of OpenSSL external methods that we subsequently implemented in TLSonSGX.
The OpenSSL library implements three data structures: \texttt{SSL\_METHOD}, \texttt{SSL\_CTX}, and \texttt{SSL}.

These data structures all contain crucial information for TLS connection security, therefore we create and confine them within the enclave.
The objects are passed by the OvS instance via an unmodified API using the external methods we implemented.
They are created, confined, and handled inside the enclave during the operation of the virtual switch, and hence discarded and not passed to ECALLs.
There is no one-to-one mapping in mbed~TLS for these three structures, hence we redefine these structures 
using mbed~TLS primitives (specifically the \texttt{mbedtls\_ssl\_config} and \texttt{mbedtls\_ssl\_context} data structures). 

The code in \texttt{stream-ssl.c} implements the interface between OvS and the OpenSSL library. 
We extended the OvS configuration script and \texttt{stream-ssl.c} with a new compilation flag, \texttt{SGX}.
If the \texttt{SGX} flag is set at compilation time, \texttt{stream-ssl.c} will use the TLSonSGX static library instead of the OpenSSL library.
Moreover, the sections of \texttt{stream-ssl.c} that load keys and certificates from the file system become redundant and are omitted.

\subsection{Application plane}
On the application plane, the solution consists of three major components: the network application, the attestation agent, and the certificate authority.

The attestation agent is a service running on the container host, setup to listen to connections from containers running on the same host, as those are the only containers able to request a quote from this host.
The attestation agent can return both a copy of the measurement list, and a quote from the TPM.
The quotes are made over the appropriate PCR registers to capture the current configuration, together with a nonce to prevent replay attacks.
Interfacing with the TPM is implemented using the TrouSerS TSS library~\cite{trousers} on Linux.
Using an attestation agent reduced the code base of the containers, since they do not have to interface directly with a TPM or Linux IMA.

Next, the CA fulfills two goals.
First, it validates the integrity of the components by validating the quote, and compares the configuration and measurement list to known good values.
Second, if the two values match, the CA signs the applications CSR.
We implemented this using the OpenSSL C library to create the signature with a pre-configured root certificate.
This root certificate is distributed to the SDN controller, allowing it to validate the certificate chain.

The final component is the container application.
Using mbed~TLS~\cite{mbedtls_sgx}, we implemented an application that supports the attestation sequence described earlier, and communicates with both the attestation agent and the CA.
Once the attestation sequence is finished, the application can connect to an SDN controller using the credentials generated and confined within the enclave.

\section{Evaluation}
\label{sec:eval}

\subsection{Testbed}
\label{subsec:testbed}
\begin{figure}[t]
	\parbox{.5\linewidth}{
		\centering
		\includegraphics[width=1\linewidth]{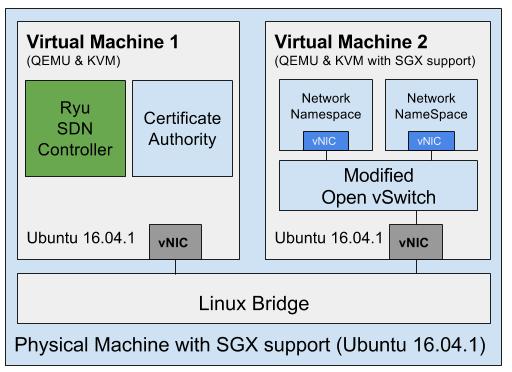}
		\caption{Testbed architecture}
		\label{fig:testbed}
	}
	\hfill
	\parbox{.5\linewidth}{
		\centering
		\includegraphics[width=1\linewidth]{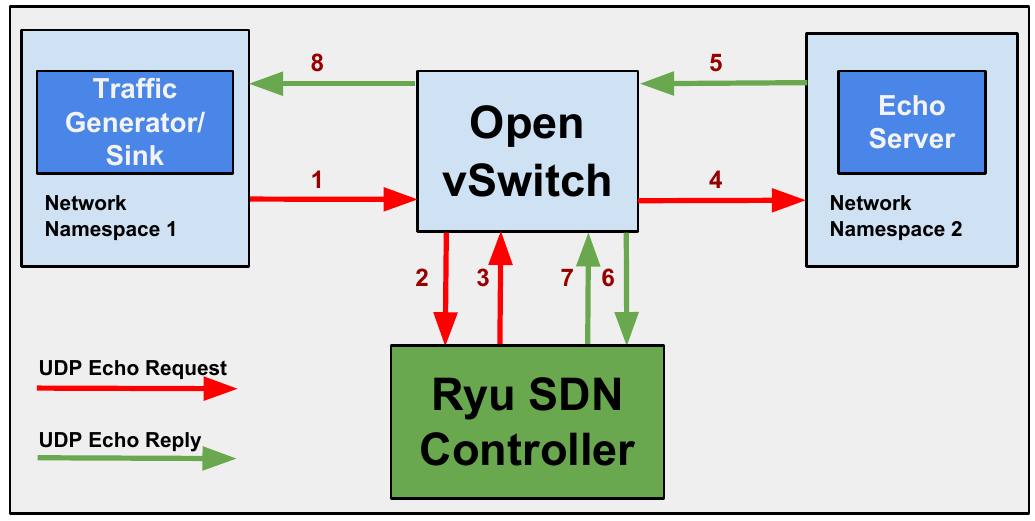}
		\caption{UDP Packet Path}
		\label{fig:Packet_Path}
	}
\end{figure}

We evaluated the solution on the testbed described below (see Figure \ref{fig:testbed}).
\paragraph{Hardware}
The host platform is a Lenovo Thinkpad T460s with a dual-core Intel\textsuperscript{\textregistered} Core\textsuperscript{TM} i7-6600U CPU clocked at 2.60GHz with SGX support.
VM\textsubscript{1} was created with 1 virtual CPU, and VM\textsubscript{2} with 2 virtual CPUs;
both VMs had 4 GB RAM, 30 GB of storage, and used virtio as vNIC.
We used Ubuntu 16.04.1 (with OvS and SGX drivers and SDKs) on both the host and VMs.
To enable the use of SGX within the VM environment, we created VM\textsubscript{2} using patched versions of QEMU and KVM provided by the SGX project\footnote{SGX Virtualization, \url{01.org/intel-software-guard-extensions/sgx-virtualization}} and Intel SGX SDK, v1.8. 

We enabled hyper-threading on the host platform, yielding 4 logical CPUs.
We pinned VM\textsubscript{1} to CPU 2 and VM\textsubscript{2} to CPUs 1 and 3 (same core).
In VM\textsubscript{2}, we pinned the virtual switch to CPU 1 and the traffic generator/sink and echo server to CPU 2, in order to reduce inter-core communication overhead~\cite{middleboxes}. 
However, due to the limited number of cores on the host (2 cores) we were unable to implement strict CPU isolation by dedicating entire cores.
In~\S\ref{subsubsec:results_latency} we discuss the potential implications of this.

\paragraph{Software}
We used OvS release 2.6.0\footnote{Commit \texttt{4b27db644a8c8e8d2640f2913cbdfa7e4b78e788}}.
In VM\textsubscript{2}, we deployed OvS binaries compiled and linked with our TLSonSGX (as explained in~\S\ref{subsec:data-plane-trust}).
We created two network namespaces, each with a port connected to the OvS instance.

The CA uses OpenSSL 1.1.0d for TLS communication with OvS and to sign the OvS and the SDN controller certificates. 
We used OpenSSL, rather than TLSonSGX for the CA implementation for two reasons: 
(1) the CA implementation is trusted according to the threat model; 
and (2) to ensure interoperability between TLSonSGX (on the client side) and OpenSSL (on the server side).

We chose the Ryu SDN open-source controller as it supports TLS communication with OpenFlow switches\footnote{See Ryu 4.9 Documentation, \url{https://ryu.readthedocs.io/en/latest/tls.html}}.
It is written in Python and is widely used in research \cite{sdn_controller_comparison} and in commercial products\footnote{See SmartSDN Controller, \url{https://osrg.github.io/ryu-book/}}.

\subsection{Evaluation Targets}
\label{sec:planned_measurements}

\paragraph{SDN Controller Program}
In the SDN model, the virtual switch forwards the first packet in a new flow to the SDN controller.
The controller replies with a flow table update, the action to be executed by the switch to handle the packet, and the packet itself.
The virtual switch handles subsequent packets in the flow according to the newly installed rule in the flow table.

To exercise the communication between the SDN controller and the virtual switch and to capture latency measurements, we designed the SDN controller as a learning L2 switch, with a MAC address to port number mapping table.
To collect measurements of the controller-induced latency, the SDN controller sends no flow updates to the virtual switch (otherwise we would get one measurement per new destination).
As a result, the virtual switch sends all the packets in the flow to the SDN controller and the controller returns the packets to the virtual switch along with the action to send the packet through the corresponding port.

\paragraph{Performance Measurements}
\label{sec:performance_management}
We are primarily interested in the latency and the time required to generate key pairs and to obtain a signed certificate from the CA. 
When it comes to latency, the choice of traffic generators was limited to those that can provide latency measurements.
Moreover, such measurements require that clocks of both traffic source and sink are synchronized (or co-exist in the same host).
Having investigated several traffic generators (qperf~\footnote{See \texttt{qperf} man page}, pktgen~\cite{olsson:2005}, moongen~\cite{moongen}, and Click~\cite{morris1999click}), we chose Click due to its flexibility and versatility.

We implemented a traffic generator and sink using the Click Modular Router.
This allows us to measure round trip latency for UDP packets of varying sizes, at a rate of 500 Packets Per Second (pps) using the Click element \newline \texttt{StoreUDPTimeSeqRecord}.
Increasing the rate beyond that results in much higher latency variance (see \S\ref{subsubsec:results_latency}).

We deployed the traffic generator and sink in network namespace \textit{(i)} and a UDP echo server in network namespace \textit{(ii)}. 
The echo server echoes the received UDP packet back to the traffic generator and sink. 
The the two network namespaces communicate through Open vSwitch, as illustrated in Figure \ref{fig:Packet_Path}.
To benchmark the performance, we replicated the measurements in a clone of VM\textsubscript{2}, using a vanilla QEMU and 
KVM, with a default Open vSwitch implementation that uses OpenSSL.

\subsection{TLSonSGX Performance Evaluation}
\paragraph{Keys and Certificate Generation Time}
\label{section:keys_time}
This measurement concerns the time from \texttt{SSL\_library\_init} invocation in the Open vSwitch until the key pairs and signed certificate are loaded to the enclave's memory.
See measurement results in Table \ref{table:key_generation}.
There is no corresponding measurement in a vanilla Open vSwitch, since keys and certificates are handled manually~\cite{ovs_ssl}. 
However, as this operation is only executed once when \texttt{ovs-vswitchd} starts, the measurements show that there is little \textit{de facto} overhead introduced by the implementation. 

\begin{table}[tb]
	\parbox{.40\linewidth}{
		\centering
		\caption{Keys and certificate generation time. 1000 measurements.}
		\label{table:key_generation}
		\begin{tabular}{|l|l|}
			\hline
			\textbf{Mean}             &     0.344 seconds        \\ \hline
			\textbf{Variance}             &     0.0488        \\ \hline
			\textbf{1st Quartile} &     0.186 seconds       \\ \hline
			\textbf{Median}       &      0.276 seconds      \\ \hline
			\textbf{3rd Quartile}       &      0.434 seconds      \\ \hline
		\end{tabular}
	}
	\hfill
	\parbox{.50\linewidth}{
		\centering
		\caption{Packet rate vs. Average CPU utilization.}
		\label{table:cpu_utilization}
		\renewcommand{\arraystretch}{}
		\begin{tabular}{|c|c|c|}
			\hline
			\textbf{Packet Rate}  &  \textbf{ OpenSSL} & \textbf{TLSonSGX}      \\ \hline
			\ 500 pps &     25\% & 61\%        \\ \hline
			\ 1000 pps            &     40\% & 78\%        \\ \hline
			\ 2000 pps &     49\% & 96\%       \\ \hline
		\end{tabular}
	}
\end{table}

\subsubsection{Packet Round Trip Latency}
\label{subsubsec:results_latency}
In this section we discuss and analyze the packet round trip latency.
The measurements do not include the key generation time;
likewise, the time to establish a TLS session is not included, as it must already be established before packets can flow.
The TLS session remains active unless one of the two ends (Open vSwitch or SDN controller) terminates the session.

\paragraph{Packet Size}
\label{sec:packet_size}
The IP packet size received by the Open vSwitch from the traffic generator is bounded by the Maximum Transmission Unit (MTU) of the network namespace port connected to the Open vSwitch (1500 bytes in our tests).
Open vSwitch encapsulates the received packet 
in an OpenFlow \texttt{Packet In} message, adding an 18 bytes header \cite{oss:2015}, that is in return encapsulated in a TLS record sent from the Open vSwitch to the SDN controller.
If the packet sent by the traffic generator is larger than the MTU, then it is fragmented and Open vSwitch handles it as two separate \texttt{Packet In} messages to the SDN controller. 

The TLS record adds a 5-byte header.
Depending on the cipher suite negotiated between the server and the client, a padding field (up to 15 bytes) is added, and the TLS record is appended with a Message Authentication Code (MAC) computed over the data.
In the handshake messages exchanged between Open vSwitch and the SDN controller in our tests, the negotiated cipher suite was \texttt{ECDHE-RSA-AES256-SHA}, which provides perfect forward secrecy through the use of an Elliptic Curve Diffie-Hellman key exchange \cite{ecc_rfc}, while the bulk encryption use 256-bit AES in CBC-mode with SHA-1 for MAC \cite{aes_rfc}.

We measure the latency for increasing packet sizes ranging from 64 bytes up to 1408 bytes (in increments of 64 bytes), including the Ethernet and IP headers (minus the Cyclic Redundancy Check).
The upper limit is set to avoid subsequent fragmentation between the Open vSwitch and the SDN controller.

\paragraph{Packet Rate Selection and CPU Utilization}
We excluded outliers with a round trip latency over 2.5 milliseconds from the captured data: 5237 outliers when testing OpenSSL and 11622 outliers when testing TLSonSGX, out of 220000 samples for each implementation.
We investigated the CPU utilization to identify the cause of the outliers and the order-of-magnitude difference in the outlier numbers between the two implementations.
In both implementations, inside the VM, the first vCPU reaches 100\% utilization due to the Click packet generation process pinned to it, even at rates lower than 500 pps (i.e., 50, 100, 200 pps). 
However, the second vCPU, where \texttt{ovs-vswitchd} process is pinned, has a higher average CPU utilization when TLSonSGX is used compared to OpenSSL (see Table \ref{table:cpu_utilization}). 
Increasing the rate beyond 500 pps leads to increasing the second vCPU's utilization and average latency.
Thus, we chose 500 pps as a suitable and optimal maximum rate for further measurements and analysis.
Using SGX causes increased CPU utilization due to the overhead of transitioning to and from the memory enclave.

\begin{figure*}[t!]
\centering
\includegraphics[width=0.88\textwidth]{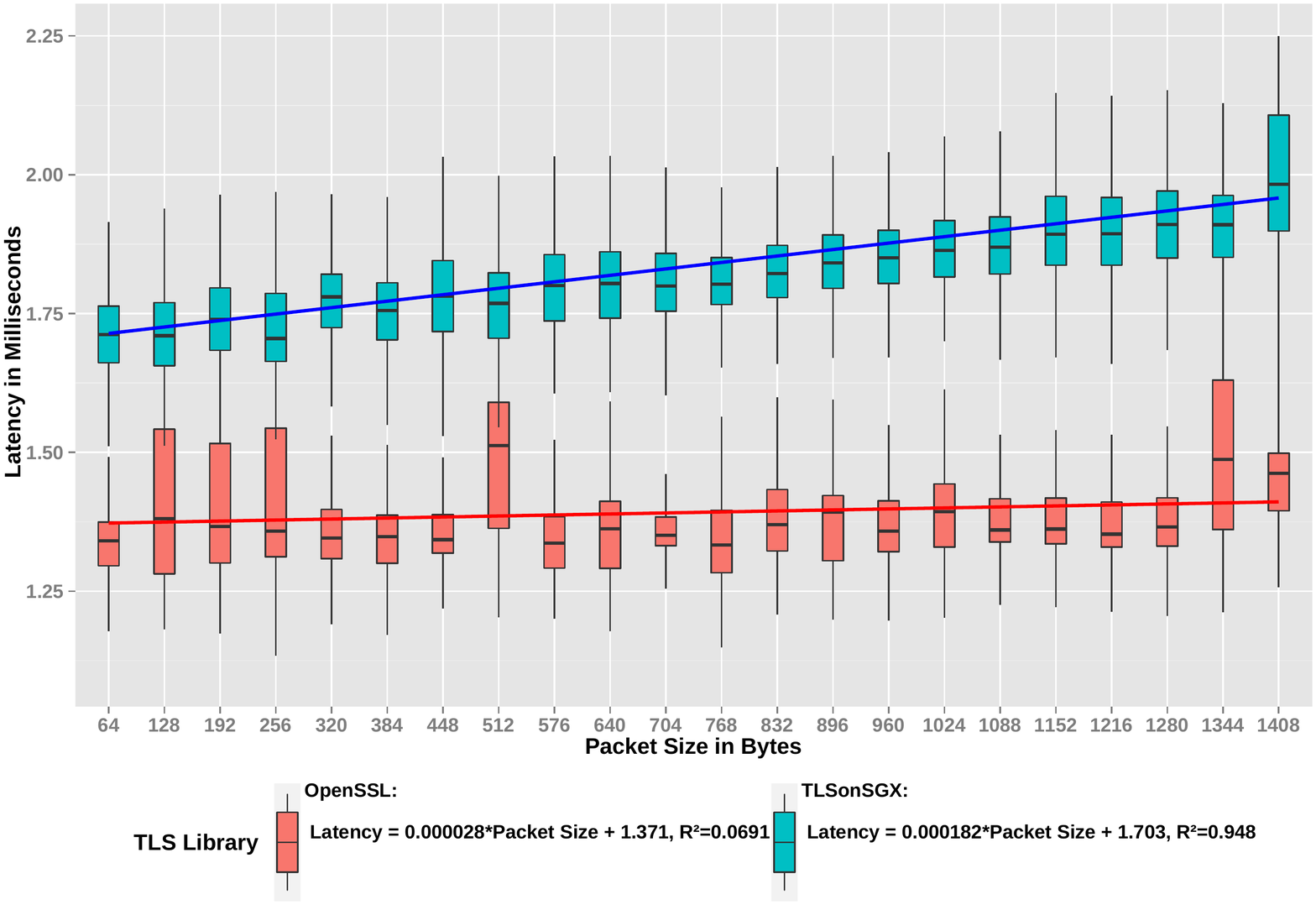}
\caption{UDP Packet Round Trip Latency vs. Packet Size}
\label{fig:tcp_latency}
\end{figure*}

\paragraph{Latency and Packet Size}
The packet round trip latency measurements are plotted in a boxplot comparing TLSonSGX with the vanilla Open vSwitch with OpenSSL when forwarding UDP packets of a range of sizes (outliers were excluded, as stated above).
Figure \ref{fig:tcp_latency} shows a plot of latency versus packet size.

Each box represents the data between first and third quartile, the thick line in the box represents the median.
The upper whisker is the minimum value between the data maximum and 3rd Quartile + 1.5*IQR, where IQR is the interquartile range.
The lower whisker is the maximum value between the data minimum and 1st Quartile - 1.5*IQR \cite{boxplot}.

A linear regression analysis of means shows that at zero byte TLSonSGX adds an overhead of 0.33 ms compared to OpenSSL.
In implementations the latency increases linearly with packet size;
we estimate this increase to 28 nanoseconds per byte for OpenSSL, and 182 nanoseconds per byte for TLSonSGX.
While the linear increase is consistent with our expectations (larger packets require more processing time),
the increase per byte is higher in TLSonSGX than in OpenSSL (154 nanoseconds per byte).
This, and the extra cost of 0.33 ms at zero byte are also expected due to the transition overhead to and from the memory enclave.

Once a packet is received at an Open vSwitch port from the network name space, \texttt{ovs-vswitchd} triggers \texttt{ecall\_ssl\_write} to encrypt and send the packet to the SDN controller,  while checking the SSL state (\texttt{ecall\_ssl\_get\_state}) before and after the write ECALL.
Since \texttt{ovs-vswitchd} uses non-blocking sockets, \texttt{ovs-vswitchd} keeps reading and returning from the socket (\texttt{ecall\_ssl\_read}), while comparing the SSL state before and after the read \newline (\texttt{ecall\_ssl\_get\_state}).
If a negative value is returned (WANT\_READ) from \texttt{ecall\_ssl\_read} then it triggers (\texttt{ecall\_ssl\_get\_error}) to retrieve the error code which indicates that the read call must be repeated and accordingly continue the loop.
If a positive value is returned, there is a response from the controller. 
The controller will respond with two packets: (1) the original packet itself; (2) the action needed by the switch to forward the packet to the second network name space.
The same flow will run during the return trip from the second network name space to the first one.

\begin{table}[t] 
	\centering
	\small
	\caption{Analysis of packet latency (all measurements are in milliseconds**)}
	\label{table:analysis}
	\setlength{\tabcolsep}{2pt}
	\begin{tabular}{rcccccccc}
		\toprule
		& & & & \multicolumn{4}{c}{\texttt{ecall\_ssl\_}} & \multirow{2}{*}{\thead{Total \\enclave \\access}} \\
		\cmidrule{5-8}
		Size (b) & 
		\thead{TLSonSGX}  &
		OpenSSL & Diff & \texttt{read} & \texttt{write} & \texttt{get\_state}* & \texttt{get\_error}* & \\
		\midrule
		64 & 1.6500 & 1.2682 & 0.3817 & 0.0047 & 0.0646 & 0.0047 & 0.0043 & 0.2966 \\
		128 & 1.6667 & 1.2722 & 0.3944 & 0.0048 & 0.0676 & 0.0047 & 0.0043 & 0.3040 \\
		256  & 1.6820 & 1.2844 & 0.3976 & 0.0049 & 0.0725 & 0.0047 & 0.0043 & 0.3146 \\
		512  & 1.6852 & 1.2955 & 0.3897 & 0.0049 & 0.0828 & 0.0047 & 0.0043 & 0.3350 \\
		1024 & 1.6963 & 1.3145 & 0.3818 & 0.0049 & 0.1022 & 0.0047 & 0.0043 & 0.3740 \\
		\bottomrule
	\end{tabular}
	
	\raggedright * \texttt{ecall\_ssl\_get\_state} and \texttt{ecall\_ssl\_get\_error} are independent of packet size. \\
	\raggedright ** Measurements captured in a different iteration than in Figure~\ref{fig:tcp_latency}.
\end{table}

To analyze and break down the time difference between OpenSSL and TLSonSGX, we traced the ECALLs indirectly called by \texttt{ovs-vswitchd} during the packet's round trip.
We measured the time consumed for each ECALL and repeated the measurement 10000 times per packet size.
Table \ref{table:analysis} lists the mean values for each of the four different ECALLs.
The last column in the table shows the sum of all ECALLs times per packet round trip. 

We noticed that the duration of \texttt{ecall\_ssl\_write} is longer (and increases with packet size) than that of other ECALLs. 
This is because \texttt{ecall\_ssl\_write} is the only ECALL that writes from a buffer with a pointer outside the enclave (unprotected memory) to the enclave memory. 
All other ECALLs do the opposite. 
According to the manual, ECALLs that pass an external pointer into the enclave are slow, since a buffer is allocated inside the enclave memory\footnote{
Pointer Handling, {Intel\textsuperscript{\textregistered} {Software Guard Extensions SDK}}, \url{https://software.intel.com/en-us/node/708975}}. 
Before copying the contents of the external buffer into the enclave memory, the content and the size of the buffer referenced by the external pointer are verified for every call to prevent overwriting enclave code or data.

Recall from the system model (consistent with a typical SDN deployment) that only the first packet in the flow is sent to the SDN controller.
As a result, crafting a small enough first packet (64 bytes) allows to optimize the latency and reduce the time to add the flow rule in the Open vSwitch flow table.

\subsection{Application Plane Evaluation}
\label{subsec:application-evaluation}

In the application plane, we are mostly interested in performance measurement regarding the attestation time.
Every time a container is launched, both the container itself and the host it is running on must be attested.
In this section, we focus on measuring the attestation time for the proposed application plane design.
There are of course other relevant performance aspects, such as time required for the actual TLS connection to the controller, but we refer to previous work for such measurements \cite{girtler-paladi-integrity:2017}.

\begin{table}[h] 
	\centering
	\caption{Attestation Time in Application Plane for various stages of the attestation sequence. Stages with execution time $< 0.010 \text{ s}$ removed.}
	\label{tbl:apptotalatttime}
	\setlength{\tabcolsep}{3pt}
	\begin{tabular}{lrrr}
		\toprule
		Stage & Mean & Variance & Median \\
		\midrule
		TPM quote              & 0.332 s & 0.000159 & 0.335 s \\
		Key generation         & 0.326 s & 0.050746 & 0.266 s \\
		CSR signing            & 0.011 s & 0.000002 & 0.010 s \\
		\midrule
		Total attestation time & 0.686 s & 0.050849 & 0.622 s \\
		\bottomrule
	\end{tabular}
\end{table}

The benchmarks were made by repeatedly launching the application which triggers the attestation. 
We ran 1000 tests, and calculated the mean and median values of the total attestation time (see results in Table~\ref{tbl:apptotalatttime}).
As seen from the table, the attestation time is well below one second in the average case.
Breaking down the execution time to various stages of the attestation, and presenting those with an execution time of $\ge 0.010 \text{ s}$, we see that the majority of the attestation time is spent in two different stages: (1) waiting for the TPM chip to generate the quote, and (2) generating the private key within the enclave. 
Stage (1) is implemented in the TPM chip itself, while stage (2) depends on the size and type of key generated. A 2048-bit RSA key was used for the measurements presented above.
We also note that our current implementation is not optimized, and it may be possible to reduce the execution time even further.


\section{Related Work}
\label{sec:related-work}

\subsection{Isolating Network Elements}
\label{subsec:effective-network-element-isolation}
Protecting the sensitive code and data of network elements is a topic of active on-going research.
Jacquin proposed an architecture that used a hardware root of trust to remotely attest the integrity of virtualization hosts in SDN infrastructure~\cite{jacquin:2015}.
Furthermore, commodity TEEs were used in case studies on securing network applications~\cite{kim:2015,shih:2016}, implemented using OpenSGX, an emulator of SGX~\cite{jain:2016}.
TruSDN is a framework for bootstrapping trust in an SDN infrastructure implemented using OpenSGX~\cite{paladi:2016b}.
It supports secure provisioning of switches in SGX enclaves, a secure communication channel between switches and SDN controller, and secure communication between endpoints in the network using session keys that are generated per flow and used only during the lifetime of the flow.
Similarly, \textit{Trusted Click}~\cite{coughlin:2017} explores the feasibility of performing network processing in SGX enclaves.

\textit{SCONE} enables operators to protect confidentiality and integrity of computation in application containers against an adversary with root access to the container host~\cite{arnautov:2016}.
SCONE achieves this by deploying containers within SGX enclaves and relies on a \texttt{libc} library ported to the SGX environment to reduce performance impact of context switches between SGX enclaves and the underlying OS, at the cost of expanding the TCB.

Our solution addresses both confidentiality of long-term credentials and session keys, as well as integrity of the network element platform. 
In particular, we enable network elements on remotely attested hosts to protect their communication with the network controller using a TLS library and credentials in a local SGX enclave.
This allows us to protect core assets with insignificant performance overhead and minimal changes to network element implementations.
Porting entire applications into SGX enclaves - as proposed in the related work above - expands the attack surface to both software vulnerabilities and side-channel attacks.
We avoid this by only porting to the enclaves a minimal TCB of the network elements.
We reduce the TCB by only confining the TLSonSGX library, credentials, and TLS session information to the enclave.

\subsection{Enrolling Network Elements}
\label{subsec:enrollment-and-key}
Incomplete or incorrect network views are an attack vector in SDN deployments~\cite{paladi:2015b}.
The Secure Network Bootstrapping Infrastructure (SNBI) protocol~\cite{snbi:2017} bootstraps secure communication channels of network elements and controllers and provisions the keys required for secure communication.
To enable connectivity to the network devices, SNBI assigns unique IPv6 addresses (based on the unique device identifier) or and bootstraps devices with the required keys.
However, the SNBI protocol is not resistant against impersonation attacks on network elements and fails to specify a protocol for software network elements with similar security features.
We address the shortcomings of SNBI by attesting the integrity of the trusted computing base of the platforms hosting network elements prior to provisioning authentication credentials;
the credentials are stored in a secure enclave and as described in~\ref{subsec:tlsonsgx-impl}, never leave the enclave.

\subsection{TLS Implementations for SGX}
\label{subsec:tls-for-sgx}
There are several known TLS libraries ported to SGX enclaves.
TaLoS~\cite{talos} terminates TLS communication inside the container enclave by providing a port of LibreSSL library into SGX and thus maintaining OpenSSL API, including APIs to set private keys and certificates from outside the enclave. 
In this paper, keys and certificates are maintained inside the enclave and no APIs are exposed to manipulate them.
Furthermore, TaLoS 
was not available at the time of writing.

Initially, mbed TLS was the only available port of a TLS library into SGX in Linux \cite{mbedtls_sgx}. 
Intel\textsuperscript{\textregistered} \cite{intel_sgxssl} and wolfSSL \cite{wolfSSL} provided a port to Linux in May 2017 and June 2017 respectively. 
However, none of these three provided an unmodified OpenSSL API that is exposed outside the enclave.
Thus, none of the TLS libraries for SGX enclaves expose the required functionality.
We implemented TLSonSGX to address the lack of usable implementations. 
TLSonSGX implements a wrapper around mbed TLS Trusted SGX library that exposes the OpenSSL APIs (that are needed for Open vSwitch TLS operations) outside the enclave.

\label{subsec:attacks-on-tls-sgx}
Popular TLS libraries with support for execution in SGX enclaves (OpenSSL, GnuTLS, mbed TLS, WolfSSL, LibreSSL) are vulnerable to Bleichenbacher attacks~\cite{bleichenbacher:1998} and a modified version padding oracle attacks~\cite{vaudenay:2002} on branch level, cache line level and page level~\cite{xiao:2017}.
Such attacks can be mitigated by using the Diffie-Hellman (DH) key exchange instead of RSA-based key exchanges and \textit{Authenticated Encryption with Associated Data} (AEAD) mode for encryption~\cite{xiao:2017}.
TLSonSGX is compatible with the mitigation suggested in~\cite{xiao:2017} and can be configured to enforce DH key exchanges and AEAD encryption mode.

\section{Limitations and future work}
\label{sec:future_work}

We implemented a prototype and tested it using one dual-core laptop and used VMs with SGX support to host the virtual switches, the SDN controller, and network namespaces (See in~\S\ref{subsec:testbed}). 
While this sufficient to demonstrate the feasiblity of TLSonSGX and compare it to OpenSSL, the platform choice 
limited possible performance measurements.
Dedicated multi-core platforms, or cloud resources, with SGX support could be used to refine the performance measurements.

The current implementation supports only one virtual switch connecting multiple VMs per physical host, as only one SSL context is created and kept inside the enclave. 
This can be improved by introducing support for multiple switches per host by extending the library to support multiple SSL contexts. 
TLSonSGX could also be extended to protect the flow table 
or OVS database content 
from tampering by storing them in the enclave. 

For keys and certificates to survive host reboots, the enclave could deploy sealing mechanisms to seal the enclave, i.e. encrypt it, export it from the enclave, and store it on the local hard disk. 
We did not prioritize this, as generating new keys and obtaining a new certificate takes approximately 0.3 seconds (See~\S\ref{section:keys_time}).


\section{Conclusion}
\label{sec:conclusion}
Protecting network elements on the data and application planes is essential for the security of SDN deployments and the network isolation between tenants.
However, both state of art network elements and the underlying platforms are vulnerable to software attacks, potentially exposing authentication credentials stored in plaintext.
To address this, we implement the TLSonSGX library that provides a secure and scalable mechanism for network elements to generate keys and obtain signed certificates, while keeping them secure within a memory enclave.
TLSonSGX confines all the TLS connections to the SDN controller within the enclave to ensure that keys, certificates, and session data remain inaccessible outside the enclave.
We complement TLSonSGX with additional mechanisms to asses the network element trustworthyness and apply the approach on both data- and application planes.

Our evaluation results show that TLSonSGX does not significantly impact the time to generate credentials and only adds an insignificant overhead when processing the first packet in each flow.
TLSonSGX reduces the TLS configuration overhead and improves the security of SDN deployments.


\subsubsection*{Acknowledgements.}
\label{sec:acknowledgements}

This research was conducted within the 5G-ENSURE and COLA projects and received  funding  from  the European Union's Horizon 2020 research and innovation programme, under grant agreements No 671562 and 731574.

\bibliographystyle{splncs04}
\bibliography{bt-bibliography}

\begin{thebibliography}{10}
\providecommand{\url}[1]{\texttt{#1}}
\providecommand{\urlprefix}{URL }
\providecommand{\doi}[1]{https://doi.org/#1}

\bibitem{anati:2013}
Anati, I., Gueron, S., Johnson, S., Scarlata, V.: {Innovative technology for
  CPU based attestation and sealing}. In: Proc. 2nd International Workshop on
  Hardware and Architectural Support for Security and Privacy. p.~10. HASP '13,
  ACM (June 2013)

\bibitem{antikainen:2014}
Antikainen, M., Aura, T., S{\"a}rel{\"a}, M.: {Spook in Your Network: Attacking
  an SDN with a Compromised OpenFlow Switch}, pp. 229--244. NordSec '14,
  Springer (October 2014)

\bibitem{sdn_controller_comparison}
Arbettu, R.K., Khondoker, R., Bayarou, K., Weber, F.: {Security analysis of
  OpenDaylight, ONOS, Rosemary and Ryu SDN controllers}. In: 2016 17th
  International Telecommunications Network Strategy and Planning Symposium
  (Networks). pp. 37--44 (Sept 2016)

\bibitem{arnautov:2016}
Arnautov, S., Trach, B., Gregor, F., Knauth, T., Martin, A., Priebe, C., Lind,
  J., Muthukumaran, D., O'Keeffe, D., Stillwell, M.L., Goltzsche, D., Eyers,
  D., Kapitza, R., Pietzuch, P., Fetzer, C.: {SCONE: Secure Linux Containers
  with Intel SGX}. In: Proc. 12th USENIX Conference on Operating Systems Design
  and Implementation. pp. 689--703. OSDI'16, USENIX (November 2016)

\bibitem{talos}
Aublin, P.L., Kelbert, F., O'Keeffe, D., Muthukumaran, D., Priebe, C., Lind,
  J., Krahn, R., Fetzer, C., et~al.: {TaLoS: Secure and Transparent TLS
  Termination inside SGX Enclaves}. Tech. Rep. 2017/5, Imperial College London
  (Mar 2017)

\bibitem{ball:2014}
Ball, T., Bj{\o}rner, N., Gember, A., Itzhaky, S., Karbyshev, A., Sagiv, M.,
  Schapira, M., Valadarsky, A.: {VeriCon: Towards Verifying Controller Programs
  in Software-defined Networks}. In: Proc. 35th ACM SIGPLAN Conference on
  Programming Language Design and Implementation. pp. 282--293. PLDI '14, ACM
  (June 2014)

\bibitem{berde:2014}
Berde, P., Gerola, M., Hart, J., Higuchi, Y., Kobayashi, M., Koide, T., Lantz,
  B., O'Connor, B., Radoslavov, P., Snow, W., Parulkar, G.: {ONOS: Towards an
  Open, Distributed SDN OS}. In: Proc. 3rd Workshop on Hot Topics in Software
  Defined Networking. pp.~1--6. HotSDN '14, ACM (August 2014)

\bibitem{bifulco:2016}
Bifulco, R., Boite, J., Bouet, M., Schneider, F.: {Improving SDN with InSPired
  Switches}. In: Proc. Symposium on SDN Research. pp. 1--12. SOSR '16, ACM
  (March 2016)

\bibitem{ecc_rfc}
Blake-Wilson, S., Bolyard, N., Gupta, V., Hawk, C., Moeller, B.: {The open
  vSwitch database management protocol}. {RFC}~4492, IETF (May 2006),
  \url{http://www.rfc-editor.org/rfc/rfc4492.txt}

\bibitem{bleichenbacher:1998}
Bleichenbacher, D.: Chosen ciphertext attacks against protocols based on the
  rsa encryption standard pkcs {\#}1. In: Proc. 18th Annual International
  Cryptology Conference. pp. 1--12. CRYPTO'98, Springer Berlin Heidelberg
  (August 1998)

\bibitem{aes_rfc}
Chown, P.: {Advanced Encryption Standard (AES) Ciphersuites for Transport Layer
  Security (TLS)}. {RFC}~3268, IETF (May 2002),
  \url{http://www.rfc-editor.org/rfc/rfc3268.txt}

\bibitem{coker:2011}
Coker, G., Guttman, J., Loscocco, P., Herzog, A., Millen, J., O'Hanlon, B.,
  Ramsdell, J., Segall, A., Sheehy, J., Sniffen, B.: {Principles of remote
  attestation}. {International Journal of Information Security}
  \textbf{10}(2),  63--81 (April 2011)

\bibitem{oss:2015}
Consortium, O.S.: {OpenFlow switch specification, v.1.5.1}. Tech. Rep. ONF
  TS-025, {Open Networking Foundation} (March 2015)

\bibitem{coughlin:2017}
Coughlin, M., Keller, E., Wustrow, E.: {Trusted Click: Overcoming Security
  Issues of NFV in the Cloud}. In: Proc. ACM International Workshop on Security
  in Software Defined Networks \& Network Function Virtualization. pp. 31--36.
  SDN-NFVSec '17, ACM (March 2017)

\bibitem{dierks:2008}
Dierks, T., Rescorla, E.: {The Transport Layer Security (TLS) Protocol Version
  1.2}. {RFC}~5246, IETF (Aug 2008),
  \url{http://www.rfc-editor.org/rfc/rfc3268.txt}

\bibitem{moongen}
Emmerich, P., Gallenm{\"u}ller, S., Raumer, D., Wohlfart, F., Carle, G.:
  Moongen: A scriptable high-speed packet generator. In: Proceedings of the
  2015 Internet Measurement Conference. pp. 275--287. IMC '15, ACM, New York,
  NY, USA (2015)

\bibitem{enns:2011}
Enns, R., Bjorklund, M., Schoenwaelder, J.: {Network configuration protocol
  (NETCONF)}. {RFC}~6241, IETF (June 2011),
  \url{http://www.rfc-editor.org/rfc/rfc6241.txt}

\bibitem{boxplot}
Frigge, M., Hoaglin, D.C., Iglewicz, B.: {Some Implementations of the Boxplot}.
  The American Statistician  \textbf{43}(1),  50--54 (1989),
  \url{http://www.jstor.org/stable/2685173}

\bibitem{girtler-paladi-integrity:2017}
Girtler, D., Paladi, N.: {Component integrity guarantees in Software-Defined
  Networking infrastructure}. In: Proc. 2017 IEEE Conf. Network Function
  Virtualization and Software Defined Networks. pp. 292--296. NFV-SDN'17
  (November 2017)

\bibitem{etsinfv:2013}
{Group Specification}: {Network Functions Virtualisation (NFV), Architectural
  Framework, v.1.1.1}. Tech. Rep. gs nfv 002, European Telecommunications
  Standards Institute (October 2013)

\bibitem{hu:2014}
Hu, H., Han, W., Ahn, G.J., Zhao, Z.: {FLOWGUARD: Building Robust Firewalls for
  Software-defined Networks}. In: Proc. 3rd Workshop on Hot Topics in Software
  Defined Networking. pp. 97--102. HotSDN '14, ACM (August 2014)

\bibitem{trousers}
{IBM Corp.}: {TrouSerS}: The open-source {TCG Software Stack},
  \url{http://trousers.sourceforge.net/}, accessed 2018-04-13

\bibitem{intel:2017}
{Intel}: {Intel 64 and IA-32 Architectures Software Developer’s Manual,
  Combined Volumes: 1, 2A, 2B, 2C, 2D, 3A, 3B, 3C, 3D and 4}. Tech. Rep.
  325462-063US, Intel Inc. (July 2017)

\bibitem{intel_sgxssl}
{Intel Corp.}: {Intel SGX SSL}, \url{https://github.com/01org/intel-sgx-ssl},
  {Accessed 2017-07-20}

\bibitem{f:2016}
Izard, R.: {Floodlight REST API}.
  \url{https://floodlight.atlassian.net/wiki/display/floodlightcontroller/Floodlight+REST+API},
  accessed: 2016-12-16

\bibitem{jacquin:2015}
Jacquin, L., Shaw, A.L., Dalton, C.: {Towards trusted software-defined networks
  using a hardware-based Integrity Measurement Architecture}. In: Proc. 1st
  IEEE Conf. Network Softwarization. pp.~1--6. NetSoft'15 (April 2015)

\bibitem{jain:2016}
Jain, P., Desai, S., Kim, S., Shih, M.W., Lee, J., Choi, C., Shin, Y., Kim, T.,
  Kang, B.B., Han, D.: {OpenSGX: An Open Platform for SGX Research}. In: Proc.
  2016 Network and Distributed System Security Symposium. NDSS '16, Internet
  Society (February 2016)

\bibitem{kim:2017}
Kim, S., Han, J., Ha, J., Kim, T., Han, D.: Enhancing security and privacy of
  tor{\textquoteright}s ecosystem by using trusted execution environments. In:
  14th USENIX Symposium on Networked Systems Design and Implementation. pp.
  145--161. NSDI '17, USENIX (2017)

\bibitem{kim:2015}
Kim, S., Shin, Y., Ha, J., Kim, T., Han, D.: {A First Step Towards Leveraging
  Commodity Trusted Execution Environments for Network Applications}. In: Proc.
  14th ACM Workshop on Hot Topics in Networks. pp. 7:1--7:7. HotNets-XIV, ACM
  (November 2015)

\bibitem{kreutz:2013}
Kreutz, D., Ramos, F., Verissimo, P.: {Towards secure and dependable
  software-defined networks}. In: Proc. 2nd ACM SIGCOMM workshop on Hot topics
  in software defined networking. pp. 55--60. HotSDN '13, ACM (August 2013)

\bibitem{mbedtls_sgx}
{mbedTLS}: {TLS for SGX: a port of mbedtls},
  \url{https://github.com/bl4ck5un/mbedtls-SGX}, {Accessed 2018-04-23}

\bibitem{mckeen:2016}
McKeen, F., Alexandrovich, I., Anati, I., Caspi, D., Johnson, S., Leslie-Hurd,
  R., Rozas, C.: {Intel Software Guard Extensions (Intel SGX) Support for
  Dynamic Memory Management Inside an Enclave}. In: Proc. 2016 Hardware and
  Architectural Support for Security and Privacy. pp. 10:1--10:9. HASP '16, ACM
  (June 2016)

\bibitem{mckeen:2013}
McKeen, F., Alexandrovich, I., Berenzon, A., Rozas, C.V., Shafi, H.,
  Shanbhogue, V., Savagaonkar, U.R.: {Innovative Instructions and Software
  Model for Isolated Execution}. In: Proc. 2nd International Workshop on
  Hardware and Architectural Support for Security and Privacy. pp. 10:1--10:1.
  HASP '13, ACM (June 2013)

\bibitem{mckeown:2008}
McKeown, N., Anderson, T., Balakrishnan, H., Parulkar, G., Peterson, L.,
  Rexford, J., Shenker, S., Turner, J.: {OpenFlow: Enabling Innovation in
  Campus Networks}. ACM SIGCOMM Computer Communication Review  \textbf{38},
  69--74 (April 2008)

\bibitem{morris1999click}
Morris, R., Kohler, E., Jannotti, J., Kaashoek, M.F.: The {Click} modular
  router. ACM Trans. Comput. Syst.  \textbf{18}(3),  263--297 (Aug 2000)

\bibitem{nyman:2014}
Nyman, T., Ekberg, J.E., Asokan, N.: {Citizen Electronic Identities Using TPM
  2.0}. In: Proc. 4th International Workshop on Trustworthy Embedded Devices.
  pp. 37--48. TrustED '14, ACM (2014)

\bibitem{olsson:2005}
Olsson, R.: Pktgen the linux packet generator. In: Proc. Linux Symposium. pp.
  11--24. Ottawa, Canada (May 2005)

\bibitem{ovs_ssl}
{Open vSwitch}: {Open vSwitch Manual},
  \url{https://github.com/openvswitch/ovs/blob/master/INSTALL.SSL.rst},
  {Accessed 2017-11-10}

\bibitem{snbi:2017}
{OpenDaylight Community}: {Secure Network Bootstrapping Infrastructure }
  (October 2017),
  \url{http://docs.opendaylight.org/en/stable-boron/user-guide/snbi-user-guide.html},
  [Online; October 2017]

\bibitem{paladi:2015b}
Paladi, N., Gehrmann, C.: Towards secure multi-tenant virtualized networks. In:
  2015 IEEE Trustcom/BigDataSE/ISPA. vol.~1, pp. 1180--1185 (Aug 2015)

\bibitem{paladi:2015}
Paladi, N.: {Towards Secure SDN Policy Management}. In: Proc. 8th International
  Conference on Utility and Cloud Computing. pp. 607--611. UCC '15 (December
  2015). \doi{10.1109/UCC.2015.106}

\bibitem{paladi:2017c}
Paladi, N.: {Trust but Verify: Trust Establishment Mechanisms in Infrastructure
  Clouds} (9 2017), {PhD Thesis, Dept. of Electrical Engineering, Lund
  University}

\bibitem{paladi:2016b}
Paladi, N., Gehrmann, C.: {TruSDN: Bootstrapping Trust in Cloud Network
  Infrastructure}. In: Proc. 12th International Conference on Security and
  Privacy in Communication Networks. pp. 104--124. SecureComm'16, Springer
  (October 2016)

\bibitem{paladi:2017b}
Paladi, N., Karlsson, L.: {Safeguarding VNF Credentials with Intel SGX}. In:
  Proceedings of the SIGCOMM Posters and Demos. pp. 144--146. SIGCOMM Posters
  and Demos '17, ACM (August 2017)

\bibitem{parno:2008}
Parno, B.: {Bootstrapping Trust in a "Trusted" Platform}. In: Proc. 3rd
  Conference on Hot Topics in Security. pp. 9:1--9:6. HOTSEC'08, USENIX (July
  2008)

\bibitem{pfaff:2013}
Pfaff, B., Davie, B.: {The open vSwitch database management protocol}.
  {RFC}~7047, IETF (December 2013),
  \url{http://www.rfc-editor.org/rfc/rfc7047.txt}

\bibitem{pfaff:2015}
Pfaff, B., Pettit, J., Koponen, T., Jackson, E., Zhou, A., Rajahalme, J.,
  Gross, J., Wang, A., Stringer, J., Shelar, P., Amidon, K., Casado, M.: {The
  Design and Implementation of Open vSwitch}. In: Proc. 12th USENIX Symposium
  on Networked Systems Design and Implementation. pp. 117--130. NSDI '15,
  USENIX (May 2015)

\bibitem{porras:2012}
Porras, P., Shin, S., Yegneswaran, V., Fong, M., Tyson, M., Gu, G.: {A security
  enforcement kernel for OpenFlow networks}. In: Proc. 1st Workshop on Hot
  topics in software defined networks. pp. 121--126. HotSDN 12, ACM (August
  2012)

\bibitem{scott:2015}
Scott-Hayward, S., Natarajan, S., Sezer, S.: {A survey of security in software
  defined networks}. IEEE Comm. Surveys \& Tutorials  \textbf{18},  623--654
  (July 2015)

\bibitem{middleboxes}
Sekar, V., Egi, N., Ratnasamy, S., Reiter, M.K., Shi, G.: Design and
  implementation of a consolidated middlebox architecture. In: {Proceedings of
  the 9th USENIX conference on Networked Systems Design and Implementation}.
  pp. 24--24. USENIX Association (2012)

\bibitem{sherwood:2010}
Sherwood, R., Chan, M., Covington, A., Gibb, G., Flajslik, M., Handigol, N.,
  Huang, T.Y., Kazemian, P., Kobayashi, M., Naous, J., Seetharaman, S.,
  Underhill, D., Yabe, T., Yap, K.K., Yiakoumis, Y., Zeng, H., Appenzeller, G.,
  Johari, R., McKeown, N., Parulkar, G.: {Carving Research Slices out of Your
  Production Networks with OpenFlow}. ACM SIGCOMM Computer Communication Review
   \textbf{40},  129--130 (January 2010)

\bibitem{shih:2016}
Shih, M.W., Kumar, M., Kim, T., Gavrilovska, A.: {S-NFV: Securing NFV States by
  Using SGX}. In: Proc. 2016 ACM International Workshop on Security in Software
  Defined Networks \& Network Function Virtualization. pp. 45--48. SDN-NFV
  Security '16, ACM (March 2016)

\bibitem{shin:2013}
Shin, S., Porras, P.A., Yegneswaran, V., Fong, M.W., Gu, G., Tyson, M.:
  {FRESCO: Modular Composable Security Services for Software-Defined Networks}.
  In: Proc. 20th Annual Network \& Distributed System Security Symposium. NDSS
  '13, Internet Society (February 2013)

\bibitem{itu-t:2016}
{Telecommunication Standardization Sector of ITU}: {Security requirements and
  reference architecture for software-defined networking}. Tech. Rep. X.1038,
  International Telecommunications Union (October 2016)

\bibitem{thimmaraju:2017}
Thimmaraju, K., Shastry, B., Fiebig, T., Hetzelt, F., Seifert, J.P., Feldmann,
  A., Schmid, S.: {The vAMP Attack: Taking Control of Cloud Systems via the
  Unified Packet Parser}. In: Proc. 2017 on Cloud Computing Security Workshop.
  pp. 11--15. CCSW '17, ACM, New York, NY, USA (2017)

\bibitem{tpm:1.2}
{Trusted Computing Group}: {TPM Main Specification Level 2 Version 1.2,
  Revision 116. Parts 1-3}. Tech. Rep. 116\_01032011, Trusted Computing Group
  Inc. (March 2011)

\bibitem{tpm:2.0}
{Trusted Computing Group}: {Trusted Platform Module Library Specification,
  Family ``2.0'', Level 00, Revision 01.16}. Tech. Rep. 120\_01102013, Trusted
  Computing Group Inc. (October 2014)

\bibitem{vaudenay:2002}
Vaudenay, S.: {Security Flaws Induced by CBC Padding - Applications to SSL,
  IPSEC, WTLS ...} In: Proc. International Conference on the Theory and
  Applications of Cryptographic Techniques: Advances in Cryptology. pp.
  534--546. {EUROCRYPT '02}, Springer-Verlag, London, UK, UK (2002)

\bibitem{wolfSSL}
{WolfSSL}: {wolfSSL with Intel SGX on Linux},
  \url{https://www.wolfssl.com/wolfSSL/Blog/Entries/2017/6/14_wolfSSL_with_Intel_SGX_on_Linux.html},
  {Accessed 2017-07-20}

\bibitem{xiao:2017}
Xiao, Y., Li, M., Chen, S., Zhang, Y.: Stacco: Differentially analyzing
  side-channel traces for detecting ssl/tls vulnerabilities in secure enclaves.
  arXiv preprint arXiv:1707.03473  (2017)

\bibitem{xing:2016}
Xing, B.C., Shanahan, M., Leslie-Hurd, R.: {Intel Software Guard Extensions
  (Intel SGX) Software Support for Dynamic Memory Allocation Inside an
  Enclave}. In: Proc. 2016 Hardware and Architectural Support for Security and
  Privacy. pp. 11:1--11:9. HASP '16, ACM (June 2016)

\bibitem{zhu:2017}
Zhu, S.Y., Scott-Hayward, S., Jacquin, L., Hill, R.: Guide to Security in SDN
  and NFV. Springer, Heidelberg, Germany, 1st edn. (2017)

\end{thebibliography}

\end{document}